# Prediction of High Curie Temperature, Large Magnetic Crystal Anisotropy in 2D Ferromagnetic $Co_2Ge_2Te_6$ Monolayer and Multilayer

Zhaoyong Guan,[†§*] Ziyuan An,[†] Yuzheng Jiang,[†] Ya Su,[ξ] Yanyan Jiang,[‡] Shuang Ni[#*]

[†]Key Laboratory of Colloid and Interface Chemistry, Ministry of Education, School of Chemistry and Chemical Engineering, Shandong University, Jinan, Shandong 250100, P. R. China

[§]Science Center for Material Creation and Energy Conversion Institute of Frontier and Interdisciplinary Science School of Chemistry and Chemical Engineering, Shandong University, Jinan, Shandong 250100, P. R. China

[ξ] School of Electrical Engineering, Shandong University, Jinan, Shandong 250100, P. R. China

[‡]Key Laboratory for Liquid-Solid Structural Evolution & Processing of Materials (Ministry of Education), School of Materials Science and Engineering, Shandong University, Jinan, Shandong, 250061, People's Republic of China

[#]Research Center of Laser Fusion, China Academy of Engineering Physics. Mianyang, Sichuan 621900, P. R. China

## ABSTRACT

The $Co_2Ge_2Te_6$ shows intrinsic ferromagnetic (FM) order, which origins from superexchange interaction between Co and Te atoms, with higher Curie temperature ($T_c$) of 161 K. $Co_2Ge_2Te_6$ monolayer (ML) is half-metal (HM), and spin-β electron is a semiconductor with gap of 1.311 eV. $Co_2Ge_2Te_6$ ML tends in-plane anisotropy (IPA), with magnetic anisotropy energy (MAE) of -10.2 meV/f.u.. $Co_2Ge_2Te_6$ ML shows good dynamical and thermal stability. Most interestingly, bilayers present ferromagnetic half-metallicity independent of the stacking orders. Notley, the multilayers ($N \geq 6$) present ferromagnetic HM, while the magnetoelectronic properties are related with the stacking patterns in thinner multilayers. Moreover, the magnetoelectronic properties are dependent on the stacking orders of bulk. The magnetic order with multilayers is determined by the super-super exchange and weak van der Waals (vdW) interaction. $Co_2Ge_2Te_6$ with intrinsic ferromagnetism, good stability of ferromagnetism and half-metallicity could help researchers to investigate its wide application in the spintronics.

## 1. INTRODUCTION

Two-dimensional (2D) intrinsic ferromagnetic materials, especial HM is urgent for the spintronics.[1, 2] All kinds of 2D materials, such as Graphene,[3] h-BN,[4] $MoS_2$,[5, 6] and stanene[7] have been successfully synthesized in recent years. However, 2D magnetic materials, especial ferromagnetic materials are rare.[8, 9] It's limited by the Mermin-Wagner theory,[10] which implies 2D magnetic materials cannot exist in the isotropic Heisenberg model at finite temperature. $CrI_3$,[11] $VSe_2$,[12] $FeGeTe_2$,[13, 14] $CrGeTe_3$ (CGT) ML[15-17], self-intercalation of 2D layered materials[18, 19] with intrinsic ferromagnetism have been successfully synthesized in the experiments. 2D magnetic materials have wide application in the condensed physics and spintronics. Therefore, 2D magnetic materials are becoming research hot.[1, 20] Ideal 2D magnetic materials are expected to have attractive properties,[2, 21, 22] such as high Curie temperature ($T_c$), large magnetic crystalline anisotropy energy (MAE) with easy magnetization axis (EA) along out of the plain, and high spin polarization. For magnetic materials, half-metallic materials are quite important,[23] whose one spin channel is insulative or semiconductive, while another channel is conductive.[23] As a result, half-metallic materials could get 100% spin-polarized current, which are highly desired in the spintronics. The perfect HM used in the spintronics is expected a high $T_c$, and the semiconductive gap should be large enough.[2] Furthermore, large MAE is urgently needed for the electronics to present half-metallicity at high temperature.[24]

Most 2D materials are semiconductors, or common metals. Graphene nanoribbon (GNR), could be transformed into HM with an external electric field.[25] The chemically functioned GNR could be also transformed into HM.[26] Furthermore, carrier,[27] defect[28] could effectively convert semiconductors into HM, but these strategies are hard to achieve in the experiments.[29, 30] Among 2D magnetic materials, only $CrI_3$[11, 31] and CGT[15] are ferromagnetic semiconductors, while $VSe_2$,[12] $Fe_3GeTe_2$,[14, 32] and $CrSe_2$[33] are spin-polarized metal with FM order. Moreover, the electronic properties of $VSe_2$ are dependent on the substrate.[12] In sum, intrinsic HM is rare in 2D materials.[21] However, 2D HM is highly expected in the spintronics.[29, 30] Therefore, we have constructed and studied half-metallic $Co_2Ge_2Te_6$ ML and multilayers with intrinsic ferromagnetism by density functional theory (DFT) and a global minimum search.

In this work, the electronic and magnetic properties of $Co_2Ge_2Te_6$ are systematically investigated by DFT. $Co_2Ge_2Te_6$ ML is an intrinsic ferromagnetic material, which origins from the superexchange interaction between Co and Te atoms, with $T_c$ of 161 K. The intrinsic ferromagnetism could be concluded by the Goodenough-Kanamori-Anderson (GKA) theory.[34-36] $Co_2Ge_2Te_6$ ML is HM with a band gap of 1.311 eV for spin-β electron, while the spin-α electron is conductive. $Co_2Ge_2Te_6$ ML and bilayer intend IPA, with MAE of -10.2, -24.659 (AA), and -24.492 (AB) meV/f.u., respectively. Besides that, $Co_2Ge_2Te_6$ bilayers retain HM with FM order, independent of the stacking orders. For the multilayers $Co_2Ge_2Te_6$ ($N \geq 6$), all layers prefer to ferromagnetically couple with other layers, and they are all HM. However, the magnetoelectric properties of bulk are determined by the stacking orders. Bulk with AA-stacking shows AFM state, while AB-stacking shows FM order. However, they are all normal spin-polarized metal. The super-super exchange and vdW interaction play a significant role in determining magnetic orders in multilayers.

## 2. COMPUTATIONAL Details.

The calculation of $Co_2Ge_2Te_6$ is using plane-wave basis Vienna Ab initio Simulation Package (VASP) code,[37] based on the DFT. The Perdew-Burke-Ernzerhof (PBE)[38] is adopted to delt with 3$d$ electron's interaction. Moreover, Co's 3$d$ electron is dealt with art of the hybrid-functional HSE06[39, 40] and GGA+U method,[41] respectively. The energies of different orders, band structures, density of states (DOS), and magnetic exchanged parameters are calculated by the art of HSE06 functional. Moreover, MAE, phonon spectra, and molecular dynamics are calculated by LDA+U method. The effective onsite Coulomb interaction parameter ($U$) and exchange interaction parameter ($J$) are set to be 7.70 and 0.70 eV, respectively. The effective $U_{eff}$ ($U_{eff} = U - J$) is 7.00 eV.[42, 43] The corresponding energies of magnetic orders and electronic properties are consistent with HSE06 functional. The vacuum space in the $z$-direction is set 16 Å. The kinetic energy cutoff is set as 300 eV, and the geometries are fully relaxed until energy and force is converged to $10^{-6}$ eV and 1 meV/Å, respectively. 6×6×1 and 9×9×1 Monkhorst-Pack grids[44] are used for geometry optimization and energy calculation (HSE06), respectively. The magnetocrystalline anisotropy (MCA) energy is calculated with an energy cutoff of 400 eV, and energy is less than $1\times10^{-8}$ eV. The spin-orbital coupling (SOC) effect is also taken into account for determining the magnetic anisotropy, and the corresponding *k*-grid is adopted 19×19×1. The *k*-grid is systematically tested, shown in Figure S1. The phonon spectra and DOS are calculated using finite displacement method as implemented in the Phonopy package.[45] A 4×4×1 cell is adopted, and total energy and Hellmann-Feynman force is converged to $10^{-8}$ eV and 1 meV/Å in the phonon spectra calculation, respectively. 6000 uniform *k*-points along high-symmetry lines are used to obtain phonon spectra. In order to confirm structural dynamical stability, *Ab initio* molecular dynamics (AIMD) simulation is also performed. The constant moles–volume–temperature (NVT) ensemble with Nosé–Hoover thermostat[46] is adopted at temperature of 300 and 500 K, respectively. The time step and total time is 1 fs and 10 ps, respectively. A larger supercell (2×2×1 cell) is adopted in the AIMD simulation, to eliminate the effect of the periodic boundary condition with smaller system size. In order to describe vdW interaction, accurate DFT-D2 method[47] is used. And the calculated distance between graphene layers is 3.25 Å,[48] which is consistent with the experimental value.

## 3. RESULTS AND DISCUSSION

**3.1. Geometry of $Co_2Ge_2Te_6$ ML.** The geometry of $Co_2Ge_2Te_6$ ML is fabricated, and confirmed by particle swarm optimization (PSO)[49] based on the crystal structure analysis, shown in Figure 1 a-c. The corresponding optimized lattice parameter is $a = b = 6.881$ Å, by fitting energy with lattice parameters, which is larger than 5.989 Å of CGT.[50] This origins that ionic radius of Co atom (65) is larger than Cr atom (62). The bond length between Co and Te atoms is 2.836 Å, while the bond length between Ge and Te atoms is 2.618 Å. The bond length between Ge and Ge atoms is 2.485 Å. The $Co_2Ge_2Te_6$ ML shows $D_{3d}$ point group, which is the same with CGT. The vertical distance between Te and Te atoms is 3.639 Å, shown in Figure 1b.

The Co atom is in the center of the octahedron, similar with Cr atom in CGT. There is 1.011 $e$ electron transfer from Co atom to Ge (0.396 $e$) and Te (0.691 $e$) atoms by the bader analysis.[51] The Co atom shows $3d^84s^1$ configuration, resulting in $Co^{1+}$ ions, as one $d$ electron is taken away. Co atom has a high-spin octahedral $d^8$ configuration, leading to a large magnetic moment (MM) of 2.044 $\mu_B$, while Ge atoms have -0.012 (0.012×2) $\mu_B$. There are six Te atoms, which have -0.059 (×2), -0.047 (×2), -0.023 (×2) $\mu_B$ MM, respectively. Each supercell has two Co atoms. Therefore, there are two kinds of orders, including FM and antiferromagnetic (AFM) orders, and the corresponding spin charge density difference is shown in Figure 1 d-e, respectively. The MM mainly localizes in Co atoms, shown in Figure 1 d-e, which is consistent with above analysis. The total MM is 4.00 $\mu_B$ for FM order, while total MM is 0.00 $\mu_B$ for AFM order. In order to describe magnetic stability, we define energy difference ($\Delta E$) between FM and AFM orders: $\Delta E = E_{AFM} - E_{FM}$. And the corresponding $\Delta E$ is 0.113 eV, which implying $Co_2Ge_2Te_6$ ML shows FM ground state.

In this section, the reason for $Co_2Ge_2Te_6$ ML showing FM order is investigated. Each Co atom is coordinated by six ligands-Te in $Co_2Ge_2Te_6$ ML, and the corresponding Te-Co-Te bond angle is 91.07°, 83.27°, 103.20°, respectively. According to the Goodenough-Kanamori-Anderson rules[34-36] of superexchange theorem, it results in FM coupling (shown in Figure 1f, g). However, there is a direct exchange interaction between Co and nearby Co atoms, which intends AFM coupling, shown in Figure 1f. As a result, the ground state is determined by the competition between superexchange and direct exchange interaction, similar to $CrI_3$[29] and CGT.[15] In $Co_2Ge_2T_6$ ML, the superexchange interaction is stronger than the direct exchange interaction. In other words, the superexchange interaction originating from the hybridization between Co-$d$ and Te-$p$ orbitals dominates the exchange interaction, shown in Figure S3 a-b. Finally, $Co_2Ge_2Te_6$ ML intends FM order.

The geometrical and magnetic properties of $Co_2Ge_2Te_6$ are investigated in the above section, and the electronic properties are usually related with the geometry. The band structure and partial density of the states (PDOS) of $Co_2Ge_2Te_6$ are calculated, shown in Figure 1 h-i. The spin-α electron channel is conductive, while the spin-β electron channel is insulative. Therefore, the $Co_2Ge_2Te_6$ is HM. The Fermi-level is partially occupied by the spin-α electrons. However, the valance band maximum locates at Γ point, while the conductance band minimum locates at K point, for the spin-β electrons. Therefore, $Co_2Ge_2Te_6$ is a semiconductor with an indirect gap of 1.311 eV for spin-β electron, shown in Figure 1h. As a result, 100% spin-polarization implies $Co_2Ge_2Te_6$ could be used as spin injection and spin transport devices.[52] Furthermore, the states near the Fermi-level are mainly contributed by the Te's *p* orbitals, while the states near the Fermi-level are partially contributed by Co's $d_{xy}$, $d_{yz}$, $d_{x^2-y^2}$ and $d_{xz}$ orbitals, shown in Figure 1b, S2a, S3a, respectively. The PDOS and integrated density of the states (IDOS) of Co atoms are shown in Figure S2 a, b, respectively.

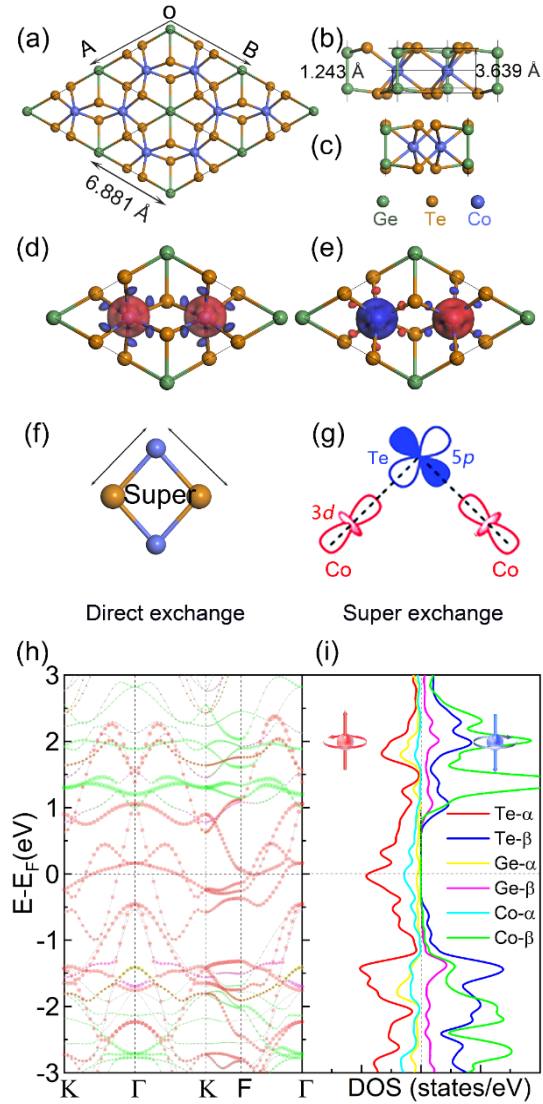

**Figure 1.** (a) Top, (b) side-1 (along *x* axis) and (c) side-2 (along *y* axis) views of optimized geometries of $Co_2Ge_2Te_6$ ML. The green, yellow and blue balls represent Ge, Te, and Co atoms, respectively. (d-e) Spin charge density difference of (c) FM and (d) AFM orders, respectively. The isovalue is 0.02 e/Å$^3$. (f) Direct and (g) superexchange interaction. (h) The atom projected band structures. (i) PDOS with FM order. The red, blue, yellow, pink, cyan, and green lines represent projected band structure and PDOS of Te-α, Te-β, Ge-α, Ge-β and Co-α and Co-β electrons, respectively. The Fermi-level is set 0 eV.

**3.2. Magnetic and Electronic Properties.** Magnetic and electronic properties of $Co_2Ge_2Te_6$ are still unknown, need to further research. The different magnetic configurations are investigated to ascertain magnetic order, shown in Figure 2a-d. Each Co atom contributes 4.0 $\mu_B$ MM, and there are eight Co atoms in the 2×2×1 cell. Therefore, there is 32.0 $\mu_B$ MM for the FM order. Moreover, three different AFM orders are considered, including AFM-zigzag (AFM-Z), AFM-stripy (AFM-S), and AFM-Néel (AFM-N) orders. For the AFM orders, four Co atoms contribute 8.0 $\mu_B$ MM, while the other four Co atoms contribute -8.0 $\mu_B$ MM. However, the MM shows different distribution. As a result, the total MM equals to 0.0 $\mu_B$, and the corresponding spin charge density difference is shown in Figure 2a-d, respectively. The energy difference is defined as the difference between AFM and FM orders. The highest energy with AFM-Z order is 0.688 eV higher than FM order, and AFM-S order has the second highest energy of 0.501 eV, shown in Figure 2b, c, respectively. Moreover, AFM-N order is 0.307 eV higher than FM order, which has the lowest energy in the AFM orders, shown in Figure 2d.

The $T_c$ is a significant parameter for ferromagnetic material, and $T_c$ of ferromagnetic materials is calculated using classic Heisenberg model Monte Carlo (MC) with the following formulas:

$$H = -J \sum_{<i,j>} S_i * S_j \quad (1)$$

$$E_{FM} = E_0 - (3J_1 + 6J_2 + 3J_3)|S|^2 \quad (2)$$

$$E_{AFM\text{-}Néel} = E_0 - (-3J_1 + 6J_2 - 3J_3)|S|^2 \quad (3)$$

$$E_{AFM\text{-}zigzag} = E_0 - (J_1 - 2J_2 - 3J_3)|S|^2 \quad (4)$$

$$E_{AFM\text{-}stripy} = E_0 - (-J_1 - 2J_2 + 3J_3)|S|^2 \quad (5)$$

Where $E_{FM}$, $E_{AFM\text{-}Néel}$, $E_{AFM\text{-}zigzag}$, and $E_{AFM\text{-}stripy}$ present energies with FM and AFM-N, AFM-Z and AFM-S orders, respectively. And $J$ and $H$ are the exchange parameter and Hamilton, respectively. And $S_i$ presents the spin operator, shown in Figure 2e. The corresponding $J_1$, $J_2$ and $J_3$ represent the nearest-, the next nearest-, and the next nearest exchanged parameter. The corresponding $J_1$, $J_2$ and $J_3$ is 3.7, 13.8, 9.0 meV, ENREF_15 respectively for $Co_2Ge_2Te_6$ ML. $J_2$ is large than $J_1$, similar phenomenon also appears in $CrB_6$-I,[53] and $Sr_2FeOsO_6$.[54] It could be concluded that long-range magnetic interactions played vital role in $Co_2Ge_2Te_6$ ML. Both nearest- and next nearest-neighbor Co atoms show FM couplings. However, $J_1$, $J_2$ and $J_3$ of CGT is 2.71, -0.058, and 0.115 meV, respectively. And the corresponding MC code is developed by Prof. Hongjun Xiang' group.[55] As a benchmark, the $T_c$ of $CrI_3$ is calculated to be 51 K,[29] which agrees well with the experimental result. A larger 80×80 cell with $1.0 \times 10^8$ loops is used to evaluate $T_c$. The 2.0 $\mu_B$ MM per Co atom drops quickly. The corresponding $T_c$ is predicted to be 161 K, which is higher than CGT (bulk, 66 K).[15]

The electronic properties of materials are usually related with the magnetic orders. The FM order is HM, while all AFM orders are spin-unpolarized metal or semiconductor, shown in Figure S4 a-d. $Co_2Ge_2Te_6$ ML under different AFM orders are spin-unpolarized semiconductor (AFM-Z) or metal (AFM-S and AFM-N). Therefore, they are different from each other. More discussion could be found in the Supporting Information.

Co is a heavier element, and the effect of SOC on the electronic properties should be considered. Therefore, the band structures with SOC are also calculated with HSE06 functional, shown in Figure S5. The band structures with EA along [100] and [001] directions are also calculated. There is a Dirac cone above the Fermi-level at Γ and K points, while the Dirac cone is absent as EA is along [001] direction, when EA is along [100] direction.

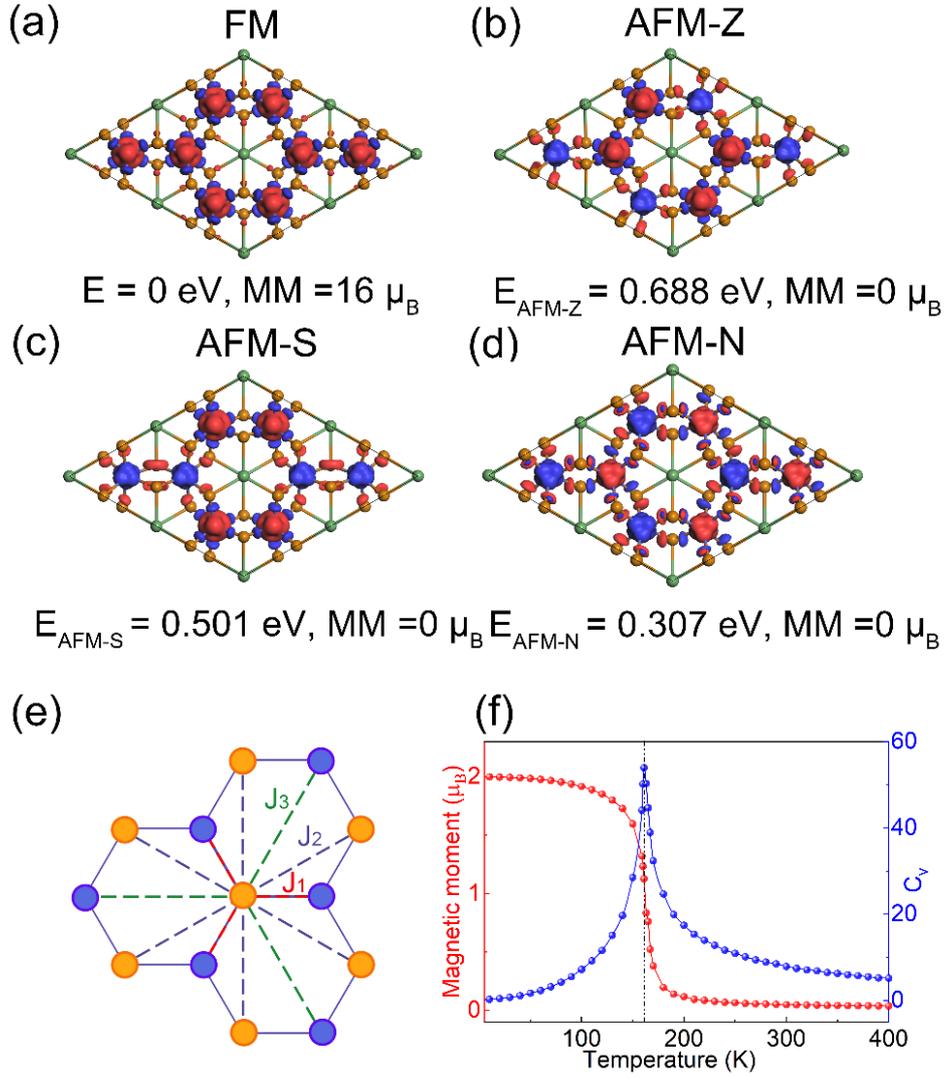

**Figure 2.** The spin charge density difference of $Co_2Ge_2Te_6$ with (a) FM, and (b) AFM-Z, (c) AFM-S, and (d) AFM-N orders. The red and blue represent spin-α and spin-β electrons. (e) Crystal structure consisting of magnetic ion Co only. Illustration of neighbor exchange interactions. $J_1$, $J_2$ and $J_3$ represent the first, second, and third in-plane nearest-neighbor spin-spin exchange interactions, respectively. (f) MM per unit cell (red) and specific heat ($C_v$) (blue) vary respect to the temperature from Heisenberg model MC simulation, respectively.

### 3.3. Magnetic Anisotropy Properties.

The MAE means electrons need energy to switch from EA (soft axis) to the other direction (hard axis). Therefore, the MAE is often used to describe the magnetic stability of materials, and ideal magnetic material is expected to have a larger MAE. In this part, the MAE and MCA are calculated using LDA+U method. The expected magnetic materials in the spintronics are expected to have higher MCA, which means electron needs more energy to overcome a higher "barrier" from EA to hard axis.[27] MCA is important for preserving the original direction of magnetic moment from heat fluctuation, especial for the HM. As $Co_2Ge_2Te_6$ has $D_{3d}$ point group, the corresponding energy ($E$) along certain direction ($\theta$, $\phi$) follows the following equations:[56]

$$\Delta E_0 = K_1 \cos^2 \theta + K_2 \cos^4 \theta + + K_3 \cos^6 \theta + K_3 \cos 3\phi \quad (6)$$
$$\Delta E_0 = E - E_{[001]} \quad (7)$$

where $E_{[001]}$ represents the energy along [001] direction. $K_1$ and $K_2$ stand for the quadratic and quartic contribution to the MAE, respectively. The energy difference $\Delta E_0$ is independent of the in-plane azimuthal angel $\phi$. Therefore, $K_3$ equals to 0, shown in Figure 3 a-b. The eq 6 is simplified into the following equation:[57]

$$\Delta E_0 = K_1 \cos^2 \theta + K_2 \cos^4 \theta + K_3 \cos^6 \theta \quad (8)$$

The $\Delta E_0$ changes as a function of polar angle $\theta$, shown in Figure 3c. And $\Delta E_0$ follows the equation: $\Delta E_0$ (meV) $= -11.74\cos^2 \theta + 4.569\cos^4 \theta - 3.038\cos^6 \theta$

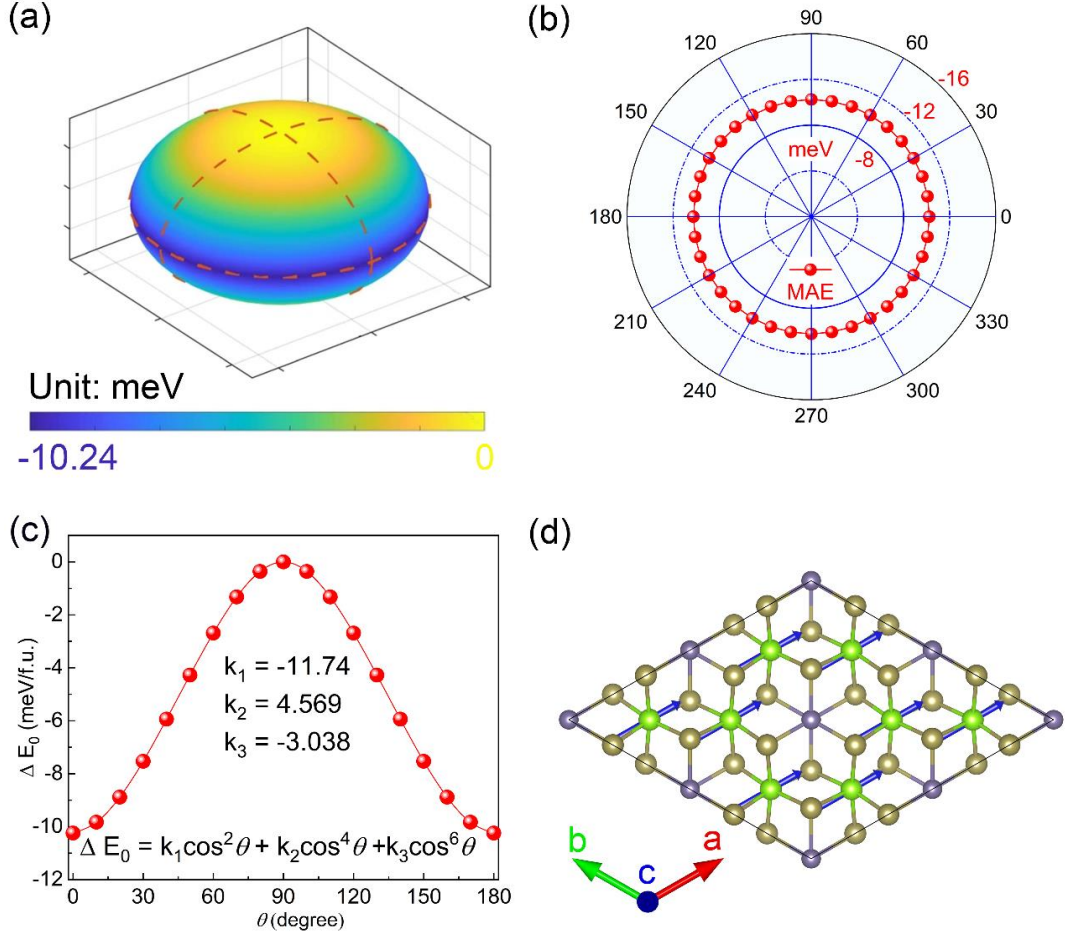

**Figure 3.** The MAE map (FM state as a reference with EA along [001]) of $Co_2Ge_2Te_6$ ML. (a) $\Delta E_0$ varies from the out-of-plane to the in-plane direction. (b) The energy indicated by the dashed lines changes with azimuthal angle $\varphi$. (c) The $\Delta E_0$ changes with polar angle $\theta$. (d) The blue arrow represents direction of EA (along [100] direction) of $Co_2Ge_2Te_6$ ML.

for $Co_2Ge_2Te_6$ ML. The MAE and MCA could be calculated using followed equations:

$$MAE = E_{[100]} - E_{[001]} \qquad (9)$$

$$MCA = E_{[100]} - E_{[001]} = MAE/S \qquad (10)$$

$E_{[001]}$ represent the energy with magnetic axis along [001] direction. $S$ is area of the supercell, and $S$ could be evaluated with this equation: $S = a^2 \sin 60°$, and $a$ is lattice parameter of unit cell. The corresponding MAE and MCA is -10.24 meV and -4.001 erg/cm$^2$, respectively. The negative MAE implies EA points to in-plane direction, shown in Figure 3 a, d. Compared with CGT (MAE = 0.5 meV),[58] the MAE of $Co_2Ge_2Te_6$ is obviously enhanced, which origins Co atom (58.93) is heavier than Cr atom (51.996). Therefore, the corresponding SOC of the former should be stronger. MAE and MCA mainly come from the contribution of SOC, similar with VSeTe.[24]

**3.6. The dynamical and thermal stability.** The dynamical stability of $Co_2Ge_2Te_6$ is confirmed via phonon dispersion curves and phonon DOS, which show no obvious imaginary phonon modes. The highest vibration frequency is 6.968 THZ, which is lower than CGT (8.364 THZ), shown in Figure 4a, S7. From Figure 4b, we can find that the contribution to the low frequency ($0 < \varepsilon < 4$ THZ) mainly comes from Te atoms. On the contrary, Ge atoms make much contribution to the high frequency ($6 < \varepsilon < 7$ THZ) parts, while Co atoms mainly make contribution to the middle frequency ($4 < \varepsilon < 7$ THZ).

The thermal stability of $Co_2Ge_2Te_6$ is also evaluated with AIMD. To examine the geometrical stability at room temperature, we also perform

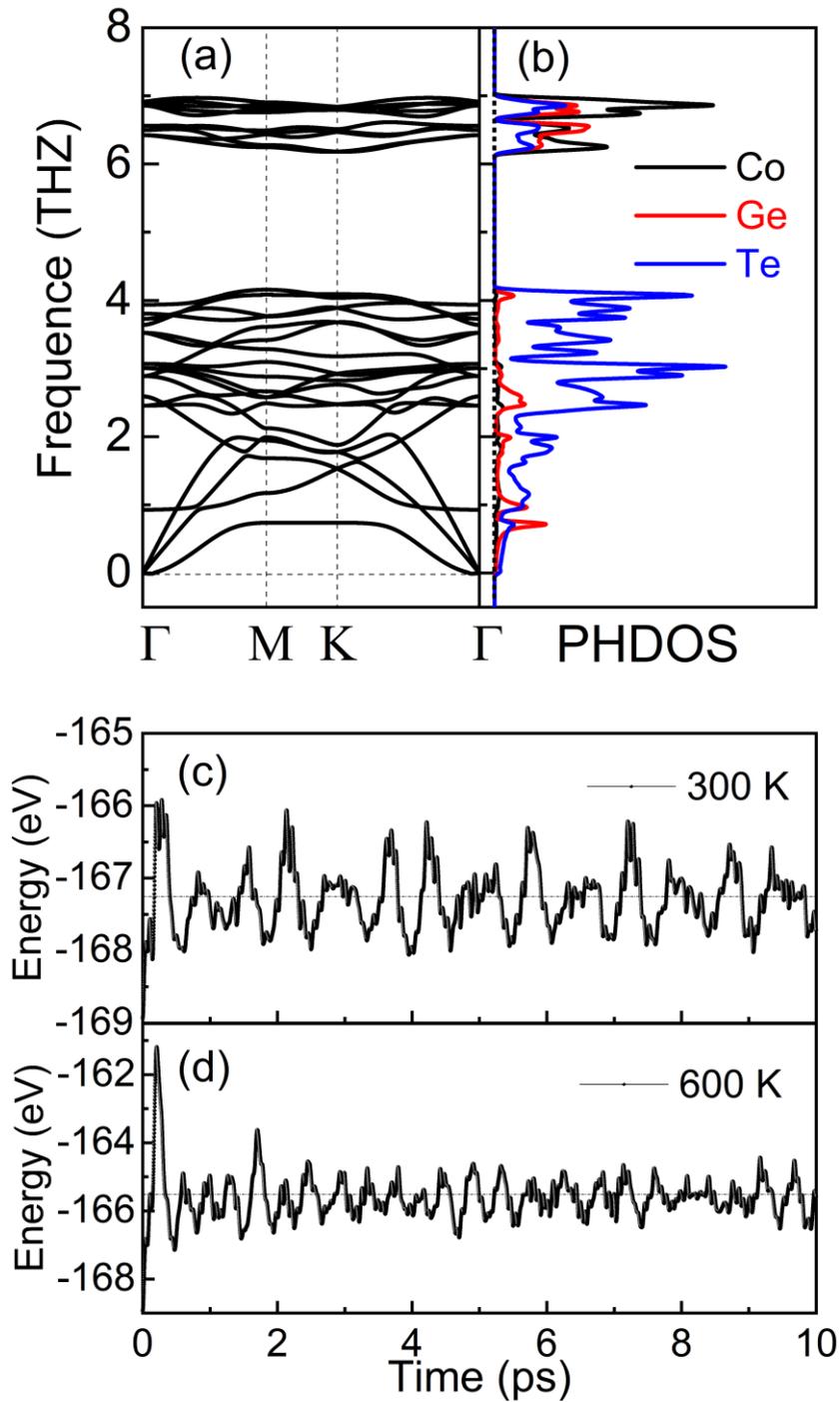

**Figure 4**. (a) The phonon band structure and (b) density of the states of $Co_2Ge_2Te_6$. The black, red, and blue lines represent partial phonon density of states of Co, Ge, and Te atoms, respectively. (c, d) The total energy and change (blue color) with the times at simulated 300 K and 600 K, respectively.

AIMD simulation at 300 and 500 K, respectively. The fluctuation in the total energies is also evaluated, and the total energies vibrate round -167.26 eV at 300 K, and -165.49 eV at 500 K, shown in Figure 4c-f. And the snapshots of the geometries also confirm the essential intact structures. No obvious structure destruction is found, so $Co_2Ge_2Te_6$ should be stable at 300 K. Moreover, geometry of $Co_2Ge_2Te_6$ ML is stable at room or higher temperature (500 K).

**3.5. Bilayer of $Co_2Ge_2Te_6$.** For the synthesized 2D materials, the two layers intend to antiferromagnetically couple with the other layer, such as $CrI_3$,[59, 60] $NiPS_3$,[61, 62] and $VSe_2$,[63] which inhibits wide application in the vdW stackings. In this part, the geometry, magnetic and electronic properties of bilayer of $Co_2Ge_2Te_6$ are systemically investigated. The bilayer of $Co_2Ge_2Te_6$ shows FM order, independent of the stacking orders. AA, AA-S-1, AA-S-2, AA-S-3, AB stackings are built, and the optimized geometries are shown in Figure 5 a-e, respectively. The vertical distance between two layers ($d_0$) is calculated, shown in Table 1. In these considering stacking orders, the AB stacking has the smallest $d_0$ of 2.999 Å, which have the lowest binding energy ($E_b$) -69.23 meV (-120.73 meV with HSE06). And AA stacking has the second smallest $d_0$ of 3.605 Å, with $E_b$ of -48.20 meV (-76.79 meV with HSE06). As for other stacking orders, such as AA-S-1, AA-S-2, AA-S-3, the corresponding $d_0$ is 3.742, 4.018, 4.100 Å, respectively. And the corresponding $E_b$ is -61.57 (-96.52), -54.78 (-91.52), -52.18 (-84.82) meV, respectively. Therefore, the AB stacking is the most stable configuration. Moreover, the $\Delta E$ is also related with the stacking orders. The AA-stacking has the lowest $\Delta E$ of -50.0 meV (-63 meV with HSE06), while AB-stacking has the second lowest $\Delta E$ of -38 meV (-56 meV by HSE06). Other $Co_2Ge_2Te_6$ with AA- stackings also have different $\Delta E$. However, $Co_2Ge_2Te_6$ with different stackings still show FM order, which is different from $CrI_3$.[64] The FM coupling between the layers comes from the super-super exchange interaction and vdW interaction between the Co atoms in the different layers. Similar trend also appears in $CrI_3$ stacking.[64]

**Table 1.** The distance ($d_0$) between two layers, binding energy ($E_b$), energy different orders ($\Delta E$) and electronic properties with different stackings are calculated by DFT+U, and HSE06 functional, respectively.

| \ | DFT+U | | | HSE06 | | Propties |
|---|---|---|---|---|---|---|
| System | $d_0$ (Å) | $E_b$ (meV) | $\Delta E$ (meV) | $\Delta E$ (meV) | $E_b$ (eV) | HM |
| AA | 3.605 | -48.20 | -50 | -63 | -76.79 | HM |
| AB | 2.999 | -69.23 | -38 | -56 | -120.73 | HM |
| AA-0.993Å | 3.742 | -61.57 | -11 | -27 | -96.52 | HM |
| AA-1.324 Å | 4.018 | -54.78 | -31 | -51 | -91.91 | HM |
| AA-6.083 Å | 4.100 | -52.18 | -19 | -34 | -84.82 | HM |

The spin charge difference of AA, AA-S-1, AA-S-2, AA-S-3, and AB stackings with FM and AFM orders are shown in Figure 5 f-g, respectively. FM order for the bilayers and multilayers is defined as followed: the Co atoms ferromagnetically couple with the inner layer, and the two layers ferromagnetically couple with the other. However, AFM order is defined as the Co atoms still ferromagnetically couple with Co atoms in the same layers, while the two layers antiferromagnetically coup with the other, shown in Figure 5 f-g. It could be found that MMs mainly localize at Co atoms, while Te and Ge atoms contribute a small part, which is similar with $Co_2Ge_2Te_6$ ML. In the most stable AB-stacking, each Co atoms has 2.433 MM under FM order, shown in Figure 5j. However, Co atoms in one layer have about 4.866 (2.433×2) $\mu_B$ MM, while Co atoms in another layer contribute -4.866 (-2.433×2) $\mu_B$ MM for the AFM order, shown in the middle inset of Figure 5j. For the AA-stacking, Co atoms contribute 4.870 $\mu_B$ MM, while Co atoms in another layer contribute -4.870 $\mu_B$ MM for the AFM order, shown in Figure 5f. For other AA-S-1 (2, 3) stackings, there's similar phenomena, and the corresponding spin charge differences are shown in Figure 5 g, h, i. The charge difference for AA and AB stackings are also calculated, shown in Figure S6. It could be found that the charges of accumulation and depletion area mainly localize in the interfaces between two layers, shown in Figure S 6a, b, respectively. Te atom gets charge, while the depletion area mainly localizes at the vacuum area between two layers.

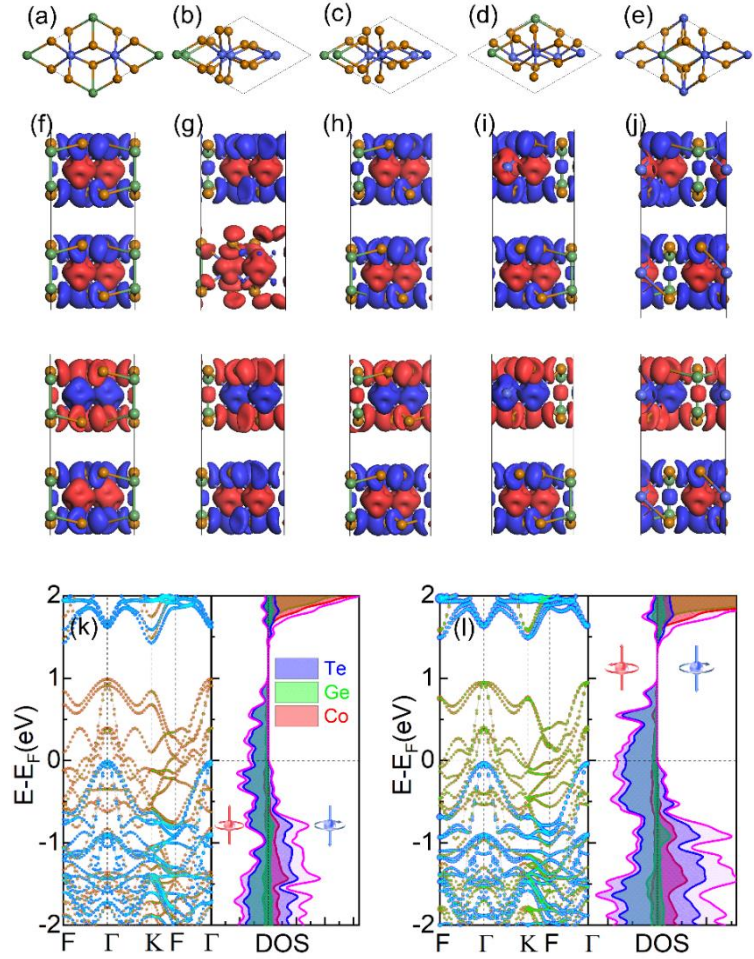

**Figure 5.** (a-e) Top views of optimized of $Co_2Ge_2Te_6$ ML with different stacking orders. The top views of ML with stackings of (a) AA, (b) AA-S-1, (c) AA-S-2, (d) AA-S-3, and (e) AB. (f-j) The spin charge density difference bilayers with different stacking orders. The red, and blue represent spin-α and spin-β electrons, respectively. The isovalue is set 0.03 e/Å$^3$. (k-l) The spin-polarized band structure and PDOS of bilayer with AA and AB stackings, respectively. The red, bule, yellow, pink, cyan and green present Te-α, Te-β, Ge-α, Ge-β, Co-α, and Co-β electrons projected band structure, respectively. The red, green, blue represent Co, Ge, Te atoms projected PDOS, respectively.

The electronic properties are usually dependent on the magnetic orders. All the considering stacking orders show FM order, and they are all HM, shown in Table 1. The layer projected band structures of AA and AB stackings are calculated with HSE06 functional, shown in Figure 5 k-l, S5 a-c, respectively. The bilayer of $Co_2Ge_2Te_6$ with AA and AB stackings show HM, and corresponding gaps of spin-β electron are 1.528 and 1.436 eV, respectively, which are larger than ML (1.311 eV). The atoms of first and second layers projected band structures are different from bilayer of MoSSe, which is caused by the quantum confinement.[48] The conduction and valance bands come from the upper and bottom layers of MoSSe, respectively.[48] While, the upper and bottom layers of $Co_2Ge_2Te_6$ show the same projected band structures, presented in Figure 5 k, l. And the states near the Fermi-level are mainly contributed by Te atoms, shown in the right columns of the Figure 5 k, l, which is the same with ML.

**3.6. Multilayers of $Co_2Ge_2Te_6$.** As number of layers (N) goes on increasing, the geometry, magnetic and electronic properties are investigated in this part. For N = 3, the ∆E is 0.347 eV (0.058 eV/Co), which implies 3L $Co_2Ge_2Te_6$ shows $Ferrim_{FM-FM-FM}$ state. $Ferrim_{FM-FM-FM}$ means the Co atom ferromagnetically couple with Co atoms with the same layer, while two layers antiferromagnetically couple with nearby layers. And the corresponding spin density is shown in the left inset of Figure 6a. The corresponding PDOS of 3L $Co_2Ge_2Te_6$ is shown in Figure 6b, which implies 3L is common spin-polarized metal. And the states at Fermi-level mainly come from Te atoms' contribution. For $N=4$, the ΔE equals to -0.382 eV (-0.048 eV/Co), which means 4L $Co_2Ge_2Te_6$ FM state (inset of Figure 6a), and the corresponding PDOS is shown in Figure 6c, which is similar with 2L $Co_2Ge_2Te_6$ with AB stacking. 4L $Co_2Ge_2Te_6$ with AB stacking is HM. The gap of spin-β electron is 1.216 eV, while spin-α electron is conducting. When $Co_2Ge_2Te_6$ is increased to 5 layers, the spin density is shown in the inset of Figure 6a, and the corresponding PDOS is shown in Figure 6d. It could be concluded that 5L $Co_2Ge_2Te_6$ is normal spin-polarized metal. When film becomes thicker, the corresponding ΔE is -0.056 ($N=6$), -0.060 ($N=7$), -0.062 ($N=8$), respectively. As N is increased to 9, 10, the corresponding ΔE are -0.074, -0.075 eV/Co, respectively, which implies FM ground state. It can be concluded that ΔE is close the bulk (-0.083 eV/Co, shown in Figure 6a), as the $Co_2Ge_2Te_6$ film becomes much thicker. The corresponding PDOS is shown in Figure 6 e, f, respectively. They are all HMs, and the similar trend also appears in $CrSe_2$ multilayer.[33] As the thickness increases, the corresponding spin-β electrons' gaps are also decreased to 1.32, 0.97, 0.66, 0.32, 0.11 eV for $N = 6-10$, respectively. It could conclude that the states near the Fermi-level are also enhanced, as $Co_2Ge_2Te_6$ multilayers become thicker.

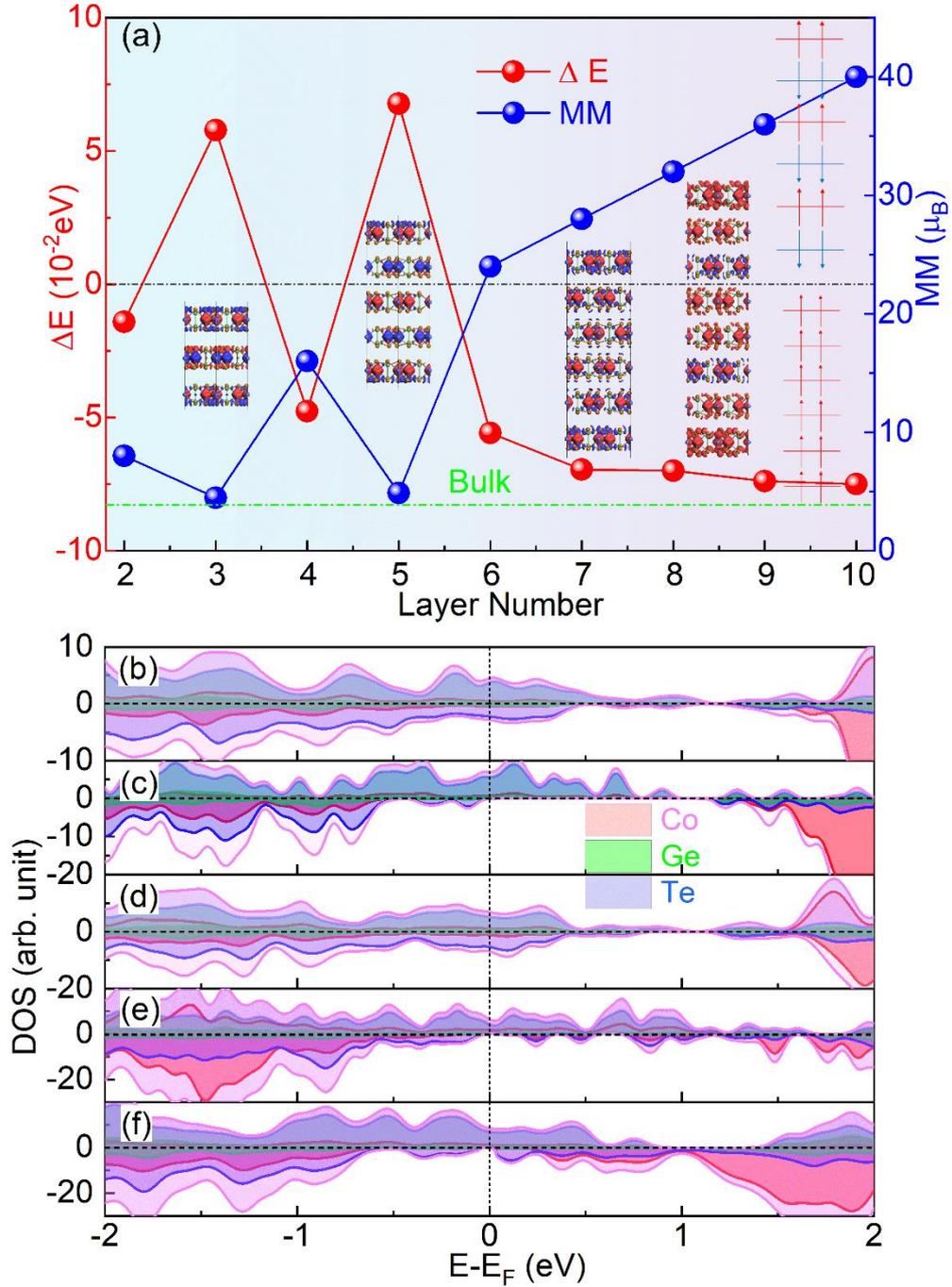

**Figure 6.** The energy difference and MM change with the layers' numbers. The inset shows the spin densities of 3L, 5L, 6L, and 8L $Co_2Ge_2Te_6$ with ground states. The PDOS of (b) 3L, (c) 4L, (d) 5L, (e) 6L, (f) 9L with ground state. The red, green, blue represent Co, Ge, Te atom's PDOS, respectively.

**3.7. Bulk of $Co_2Ge_2Te_6$.** As the thickness is further increased, the $Co_2Ge_2Te_6$ could form bulk. According to the stacking orders of the bilayer, there should be two different stacking orders: bulk-AA, bulk-AB. The corresponding geometry, magnetic and electronic properties are shown in Figure 7 a-h, 8 a-g, respectively. For the bulk-AA and AB stackings, they still have $D_{3d}$ point group, and the corresponding lattice parameters are 7.191 (Figure 7a), and 6.991 (Figure 8a) Å, respectively, which are little larger than ML. The vertical distance between Te atoms is 2.76 Å, shown in Figure 8b, which is smaller than the ML. The bulk-AA shows AFM-Z order, while bulk-AB shows FM order, shown in Figure7, and Figure 8, respectively. For AA stacking, two Co atoms intend to antiferromagnetically couple with each other, shown in the right inset of Figure 7a, S8. And the MMs mainly localize in the Co atoms, shown in Figure 7 c-f, S8 a, b, similar with ML (Figure 1 d, e). The AFM order has lower energy of -0.331 eV than FM order for the 1×1×1 cell. In order to establish magnetic ground state, the AFM-N, FM, AFM-ST, AFM-Z orders are considered, shown in Figure 7 c-f, respectively. And FM order has the highest energy, with the largest MM of 16 $\mu_B$, while the AFM-N

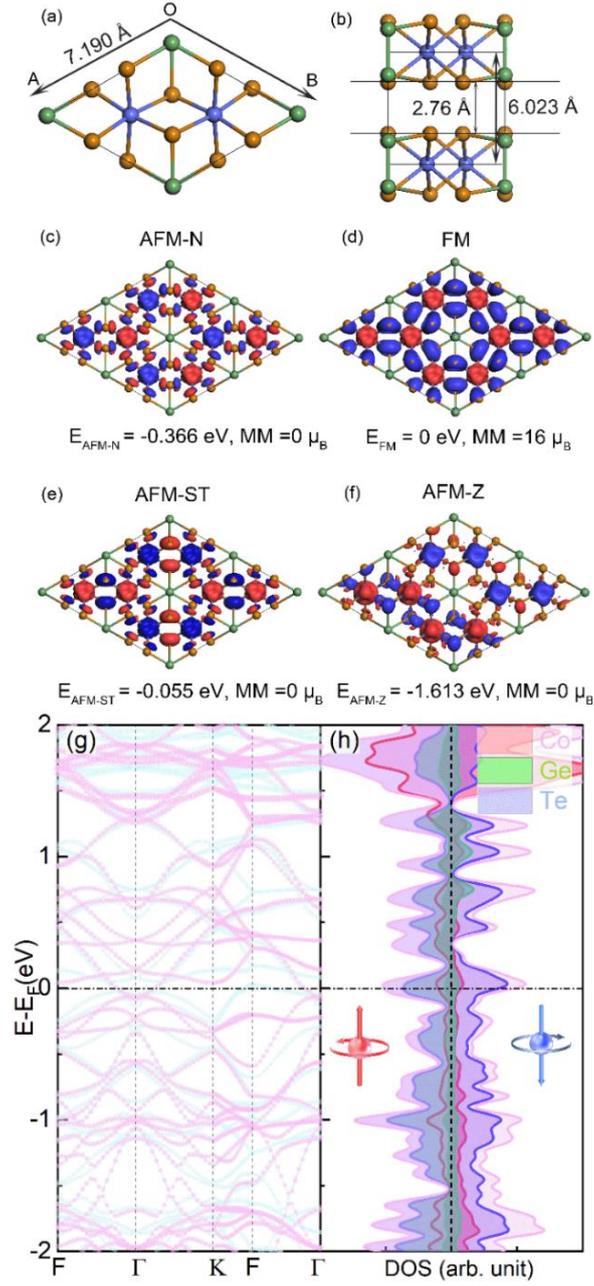

**Figure 7.** (a) Top, (b) side (along *y* axis) views of optimized geometries of bulk-AA of $Co_2Ge_2Te_6$. (d-e) Spin charge densities difference of (c) AFM-N, (d) FM, (e) AFM-ST and (f) AFM-Z orders of ML, respectively. The isovalue is 0.026 e/Å$^3$. (h) The atom projected band structures. (i) PDOS with FM order. The red, green, blue lines represent PDOS of Co, Ge, and Te electrons, respectively. The Fermi-level is set 0 eV.

order has lower energy of -0.366 eV than FM order, shown in Figure 7c. The AFM-Z order has the lowest energy (-1.613 eV), shown in Figure 7f. And the corresponding spin-polarized band structure and PDOS are shown in Figure 7 g, h, respectively. The bulk-AA order is spin-polarized metal, shown in Figure 7g. However, the $Co_2Ge_2Te_6$ ML with AFM-Z order is spin unpolarized semiconductor, shown in Figure S4b. It should be caused by the enhanced interaction between two layers, as the vertical distance between Co atoms is obviously decreased for the bulk, shown in Figure 8b. The states at the Fermi-level are mainly contributed by Te atoms, shown in Figure 8h, which is similar with ML. Compared with bulk-AA, bulk-AB has lower energy. Therefore, the bulk of AB stacking (Figure 8a) is most stable configure. Moreover, the vertical distance of Te atoms of bulk with AB stacking is 2.940 Å, shown in Figure 8b. For the 1×1×1 cell, the FM order has lower energy than AFM orders. All magnetic orders including these magnetic configurations: $FM_{FM-FM}$, $AFM_{AFM-AFM}$, and $AFM_{FM-FM}$ orders, are shown in Figure 8 a, b, c, respectively. The subscript FM-FM represents each Co atom in the same layer ferromagnetically couple with each other, shown in Figure 8c. Two layers ferromagnetically couple with each other, which is defined as $FM_{FM-FM}$ order. The $AFM_{AFM-AFM}$ (Figure 9d), and $AFM_{FM-FM}$ (Figure 8e) orders have higher energy of 0.184, and 0.331 eV than $FM_{FM-FM}$ order (Figure 8c), respectively. For the $FM_{FM-FM}$ order, Co atoms have 2.371(1$^{st}$ L), 2.371 (2$^{nd}$ L), 2.416 (2$^{nd}$ L), 2.416 (1$^{st}$

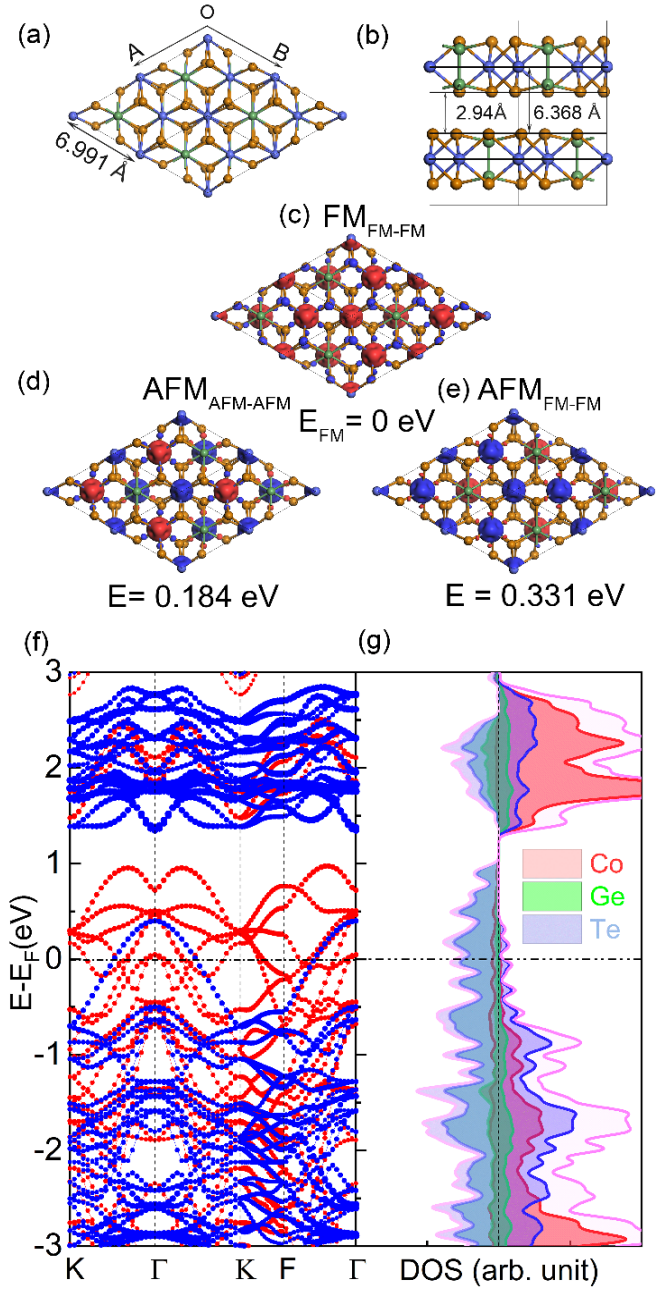

Figure 8. The optimized geometry of $Co_2Ge_2Te_6$ bulk-AB stacking with (a) top, and (b) side views of bulk-AB. The spin densities of (c) $FM_{FM-FM}$, (d) $AFM_{AFM-AFM}$, and (e) $AFM_{FM-FM}$ orders, and isovalue is set 0.026 e/Å$^3$. (f) The spin-polarized band structure and (g) PDOS of bulk-AB with $FM_{FM-FM}$ order. The red and blue represent spin-α and spin-β electrons, respectively.

L) μ$_B$, respectively, shown in Figure 8c. For AFM$_{FM-FM}$ order, Co atoms have 2.353 (1$^{st}$ L), 2.451 (1$^{st}$ L), -2.353 (2$^{nd}$ L), -2.451 (2$^{nd}$ L) μ$_B$, respectively, shown in Figure 8d. However, Co atoms have 2.359 (1$^{st}$ L), -2.422 (1$^{st}$ L), -2.359 (2$^{nd}$ L), 2.422 (2$^{nd}$ L) μ$_B$, respectively, shown in Figure 8e. Compared with HM of ML, the original states fully occupied by the spin-β electrons are shifted upward as the interlayer interaction enhances. As a result, the fully occupied states are transformed into partially occupied states, which origins from Te's contribution, shown in Figure 8g. Therefore, bulk bulk-AB is spin-polarized metal, shown in Figure 8 f, g, which should be caused by the enhanced super-super exchange interaction between the interlayer. In a conclusion, the magnetic orders are related with the stacking orders, and bulk-AA shows AFM state, while bulk-AB shows FM state, respectively.

**3.6. Magnetocrystalline Anisotropy.** In order to clarify the atomic orbital contribution to the MAE, the tight-binding and second-order perturbation theory are adopted in the calculating MAE. According to the canonical formulation,[65] MAE of each atom could be calculated, using this equation:

$$MAE_i = \left[\int E_f(E-E_F)[n_i^{[100]}(E) - n_i^{[001]}(E)]\right] \quad (13)$$

where $MAE_i$ represents the MAE of $i$th atom. $n_i^{[100]}(E)$ and $n_i^{[001]}(E)$ are the DOS of the $i$th atom with EA along [100] and [001] directions, respectively. Co$_2$Ge$_2$Te$_6$ ML has $D_{3d}$ group. Therefore, the energies with EA along [100] and [010] directions are the same.[27] So only [100] direction is considered here. Moreover, total MAE could be rewritten as the sum of $MAE_i$: $MAE_{tot} = \sum_i MAE_i$. According to the second-order perturbation theory,[66] MAE could be gotten by the sum of the following terms:

$$\Delta E^{--} = E_x^{--} - E_z^{--} = \xi^2 \sum_{o^+,u^-} (|<o^-|L_z|u^-＞|^2 - |<o^-|L_x|u^-＞|^2)/(E_u^- - E_o^-) \quad (14)$$

$$\Delta E^{-+} = E_x^{+-} - E_z^{+-} = \xi^2 \sum_{o^+,u^-} (|<o^+|L_z|u^-＞|^2 - |<o^+|L_x|u^-＞|^2)/(E_u^- - E_o^-) \quad (15)$$

where + and − represent spin-α and spin-β states, and $\xi$, $L_x$, $L_z$ are the SOC constant, angular momentum operators along [100] and [001] directions, respectively. $u$, and $o$ represent unoccupied and occupied states. And $E_o$, $E_u$ represent energies of occupied and unoccupied states, respectively. MAE is mainly contributed by the spin-orbital matrix elements and energy difference. According to the eq 13, the MAE is related with the intensity of DOS near the Fermi-level. The matrix element differences $|<o^-|L_z|u^-＞|^2 - |<o^-|L_x|u^-＞|^2$ and $|<o^+|L_z|u^-＞|^2 - |<o^+|L_x|u^-＞|^2$ for $d$ and $p$ orbitals are calculated, shown in Table 2 and Table 3, respectively. To further interpret MAE changes with number of layers, the atom-orbital-resolved MAE is also analyzed, shown in Figure 9 a-i. And it can be found that MAE partially come from Co (Figure 9 a-c) and Ge atoms' contribution (Figure 9 d-f), while it mainly comes from Te atoms' contribution (Figure 9 g-i). The orbital-resolved MAE of ML is shown in Figure 9 a, d, g, respectively. The total MAE is -10.24 meV/f.u., and Te atoms contribute -9.94 meV. The hybridization between Co's $d_{yz}$ and $d_{z^2}$, $d_{xy}$ and $d_{x^2-y^2}$ orbitals makes negative contribution to MAE (-0.22, -0.15 meV), which corresponds to the matrix differences -3 and -4 for $d$ orbitals, respectively, shown in Table 2. The hybridization between Co atoms' $d_{xz}$ and $d_{yz}$ orbitals makes positive contribution (0.39 meV) to MAE, which corresponds to the matrix differences +1 for $d$ orbitals. Ge's contribution to MAE could be negligible, compared with Te atoms.

**Table 2.** The matrix differences for $d$ orbitals between magnetization along [001] and [100] directions in eq 14 and eq 15.

| $u^-$ | $o^+$ | | | | | $o^-$ | | | | |
|---|---|---|---|---|---|---|---|---|---|---|
| | $d_{xy}$ | $d_{yz}$ | $d_{z^2}$ | $d_{xz}$ | $d_{x^2-y^2}$ | $d_{xy}$ | $d_{yz}$ | $d_{z^2}$ | $d_{xz}$ | $d_{x^2-y^2}$ |
| $d_{xy}$ | 0 | 0 | 0 | 1 | -4 | 0 | 0 | 0 | -1 | 4 |
| $d_{yz}$ | 0 | 0 | 3 | -1 | 1 | 0 | 0 | -3 | 1 | -1 |
| $d_{z^2}$ | 0 | 3 | 0 | 0 | 0 | 0 | -3 | 0 | 0 | 0 |
| $d_{xz}$ | 1 | -1 | 0 | 0 | 0 | -1 | 1 | 0 | 0 | 0 |
| $d_{x^2-y^2}$ | -4 | 1 | 0 | 0 | 0 | 4 | -1 | 0 | 0 | 0 |

When two layers are stacked with AA and AB patterns, the orbital-projected MAE is also calculated, shown in Figure 9 b, e, h (AA) and Figure 9 c, f, i (AB), respectively. The total MAEs for AA and AB stackings are -24.659, -24.492 meV, respectively, which are about two times of $Co_2Ge_2Te_6$ ML. The Te atom contributes -24.83, and -23.05 meV to total MAE, while Co and Ge atoms contribute about 0.54, -0.38, 0.47, and -0.32 meV for AA and AB stackings, respectively. Therefore, the atomic hybridization between Te atomic spin-β occupied $p_y$ and $p_z$ orbitals dominates in-plane magnetic anisotropy (IMA) (-13.33 meV), which corresponds to the matrix differences -1 for *p* orbitals. While the hybridization between occupied spin-β $p_z$ orbitals and unoccupied spin-β $p_x$ orbitals makes contribution to PMA (1.63 meV), which corresponds to the matrix 1 for *p* orbitals, shown in Table 3.

**Table 3.** The matrix differences for *p* orbitals between EA along [001] and [100] directions in eq 14 and eq 15.

|  | $o^+$ | | | $o^-$ | | |
|---|---|---|---|---|---|---|
| $u^-$ | $p_y$ | $p_z$ | $p_x$ | $p_y$ | $p_z$ | $p_x$ |
| $p_y$ | 0 | 1 | -1 | 0 | -1 | 1 |
| $p_z$ | 1 | 0 | 0 | -1 | 0 | 0 |
| $p_x$ | -1 | 0 | 0 | 1 | 0 | 0 |

dominates MAE for AA and AB stackings, which is similar with ML. Moreover, the interaction between $p_y$ and $p_z$ orbitals contribute -25.90 and -23.19 meV for AA and AB stackings, respectively, shown in Figure 9 h, i, respectively. And more detail could be found in Figure 9 a-h.

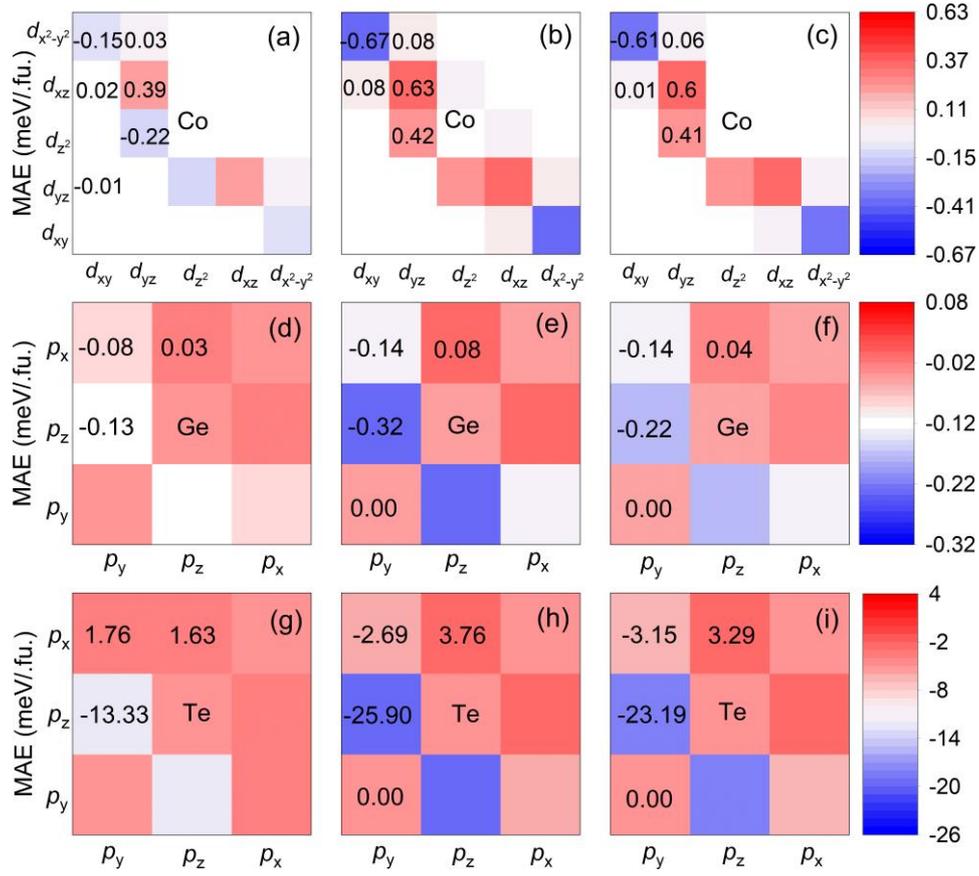

**Figure 9.** Orbital-resolved MAE of $Co_2Ge_2Te_6$ ML, bilayer with AA and AB stackings, respectively. The orbital-resolved MAE of $Co_2Ge_2Te_6$ (a, d, g) ML, (b, e, h) bilayer with AA, and (c, f, i) AB stackings, respectively.

## 3. CONCLUSIONS

In summary, we have predicted and investigated magnetic and electronic properties of $Co_2Ge_2Te_6$ ML with PSO and DFT method. We have found $Co_2Ge_2Te_6$ ML shows intrinsic ferromagnetism, which comes from the superexchange interaction between Co and Te atoms, and the corresponding bond angle is close to 90˚. $Co_2Ge_2Te_6$ ML have higher $T_c$ of 161 K. $Co_2Ge_2Te_6$ is HM with gap of 1.311 eV for spin-β electrons. The corresponding $J_1$, $J_2$ and $J_3$ of $Co_2Ge_2Te_6$ ML is 3.7, 13.8, and 9.0 meV, respectively. $Co_2Ge_2Te_6$ ML shows IMA, and corresponding MAE is -10.24 meV/f.u.. $Co_2Ge_2Te_6$ ML shows good dynamical and thermal stability. $Co_2Ge_2Te_6$ bilayer shows robust ferromagnetism and half-metallicity, independent of the stacking orders. All the layers ferromagnetically couple with other layers ($N \geq 6$), while the layer antiferromagnetically couple with other nearby layers for thinner odd layers ($N = 3, 5$). All even and thick odd multilayers are HM, while other multilayers are spin-polarized metal. The magnetoelectronic properties are dependent on the stacking orders for bulk. Bulk-AA shows AFM order, while bulk-AB shows FM order. However, they are all spin-polarized normal metal. The super-super exchange interaction and vdW interaction play a key role in the multilayers. Our work represents robust ferromagnetic half-metallic $Co_2Ge_2Te_6$ with high $T_c$, large MAE, making it a candidate for the new magnetoelectronics.

**Author Information**

Corresponding Author

*E-mail: zyguan@sdu.edu.cn; Tel: +86-0531-88363179; Fax: +86-0531-88363179

nishuang@163.com; Tel: +86-0531-88363179; Fax: +86-0531-88363179

**Acknowledgements**

We thank Prof. Wenhui Duan, and Xingxing Li for discussion of evaluation of Curie temperature. We thank Prof. Jun Hu and Jinlong Yang for discussion of MCA. This work was supported by the financial support from the Natural Science Foundation of China (Grant No. 11904203), and the Fundamental Research Funds of Shandong University (Grant No. 2019GN065). The computational resources from Shanghai Supercomputer Center. The scientific calculation in this paper have been performed on the HPC Cloud Platform of Shandong University. The authors are grateful to Beijing PARATERA Tech Corp., Ltd. for the computation resource in the National Supercomputer Center of Guangzhou. The authors are grateful to Tencent Quantum Laboratory for the computation resource.

**Conflict of Interest:** The authors declare no competing financial interest.